\documentclass[
    ,final            
  ]
  {aipprocmy}

\layoutstyle{6x9}

\begin{document}

\title{Explosions inside Ejecta and Most Luminous Supernovae}

\classification{26.30.-k, 26.30.Ef, 97.60.Bw
}
\keywords{supernovae, explosion mechanism}

\author{S.I.Blinnikov}{
  address={Institute for Theoretical and  Experimental Physics (ITEP),
Moscow, 117218, Russia}
  ,altaddress={RESCEU, School of Science, \\
Tokyo University, Tokyo 113-0033, Japan} 
}

\begin{abstract}
The extremely luminous supernova SN2006gy
is explained in the same way as other SNIIn events: light is produced by a radiative shock
propagating in a dense circumstellar envelope formed by a previous weak explosion.
The  problems in the theory and observations of multiple-explosion SNe~IIn are briefly reviewed.
\end{abstract}

\maketitle


\section{Introduction}

The discovery of SN2006gy \cite{Ofek,Smith}
demonstrates that some  supernova (SN) events produce 10 or even 100 times more visible photons
than other, already powerful explosions.
The anomalously high power of the emission of SN2006gy demands an explanation.

SN2006gy is of  SNIIn type and it revived interest in SN models where light is produced
by a long living radiative shock which  propagates in a dense circumstellar envelope.

I discuss problems in the  theory of  SNIIn events and in the physics of supercritical radiative shocks.
The powerful visible light of SN2006gy can be easily explained by a radiative shock born
due to a collision of SN ejecta with a dense cloud formed by a  weak explosion some
years before the SN.
Strong X-ray emission of SNIIn near maximum light may be absent since it is absorbed by
the dense cloud or not produced at all in the radiation-dominated shock.

\section{Supernova types}

The models with a long living radiative shock running in a dense circumstellar envelope
were invoked earlier \cite{GN:1986,ChugaiBlinn} to explain  the unusual properties of other
powerful supernovae with narrow emission lines in their spectra, SNIIn.
SN2006gy also belongs to the same SNIIn class.

A simple diagram below illustrates the relation between different astronomical types
of supernovae which are classified purely on the appearance of their spectra near maximum light,
irrespective of the underlying physics.
For example, we believe that SNe~II are born when a giant star with an H-rich atmosphere
has a powerful explosion in its core.
This explosion may be  a result  of a catastrophic collapse of the stellar core.
The details of the explosion of core-collapsing SNe are still not clear,
in spite of  many important results obtained by workers in this field.
\begin{center}
\unitlength 1mm 
\linethickness{1.4pt}
\ifx\plotpoint\undefined\newsavebox{\plotpoint}\fi 
\begin{picture}(44.25,48.5)(0,10)
\put(9.25,48.5){\line(0,-1){29}}
\put(9.25,19.5){\line(1,0){35}}
\multiput(44.25,19.5)(-.0409883721,.0337209302){860}{\line(-1,0){.0409883721}}
\put(-1,49){\makebox(0,0)[cc]{\bf Hydrogen}}
\put(18,49){\makebox(0,0)[cc]{\it type II}}
\put(2.25,16){\makebox(0,0)[cc]{\bf Hydrogen}}
\put(2.25,10){\makebox(0,0)[cc]{\rm Narrow Lines, {\it type IIn}}}
\put(56,16){\makebox(0,0)[cc]{\bf No Hydrogen}}
\put(56,10){\makebox(0,0)[cc]{\it type Ia, Ib/c}}
\end{picture}
\end{center}
Taking into account the physics of SN explosion we can produce a couple of other,
a bit different, diagrams for SN types.
I want to subdivide the SN physics into two groups of problems.
First, here is a diagram for the mechanism of explosion.
\begin{center}
\unitlength 1mm 
\linethickness{1.4pt}
\ifx\plotpoint\undefined\newsavebox{\plotpoint}\fi 
\begin{picture}(44.25,48.5)(0,10)
\put(9.25,48.5){\line(0,-1){29}}
\put(9.25,19.5){\line(1,0){35}}
\multiput(44.25,19.5)(-.80,.68){44}{\line(-1,0){.6}}
\put(-5,49){\makebox(0,0)[cc]{\bf Core Collapse}}
\put(21,49){\makebox(0,0)[cc]{\it  type II, Ib/c}}
\put(2.25,16){\makebox(0,0)[cc]{\bf Thermonuclear}}
\put(-2.25,10){\makebox(0,0)[cc]{\rm Nondegenerate, Pair Instability, {\it type IIn}}}
\put(56,16){\makebox(0,0)[cc]{\bf Thermonuclear}}
\put(56,10){\makebox(0,0)[cc]{\rm Degenerate, {\it type Ia}}}
\end{picture}
\end{center}
And the next diagram illustrates main ways to produce light during supernova explosions.
Here $S$ is entropy, and `heating' is entropy production.
\begin{center}
\unitlength 1mm 
\linethickness{1.4pt}
\ifx\plotpoint\undefined\newsavebox{\plotpoint}\fi 
\begin{picture}(44.25,48.5)(0,10)
\put(9.25,48.5){\line(0,-1){29}}
\put(9.25,19.5){\line(1,0){35}}
\multiput(44.25,19.5)(-.0409883721,.0337209302){860}{\line(-1,0){.0409883721}}
\put(-6,49){\makebox(0,0)[cc]{\bf Cooling}}
\put(-6,43){\makebox(0,0)[cc]{\bf no source of $S$}}
\put(21,49){\makebox(0,0)[cc]{\it type II}}
\put(2.25,16){\makebox(0,0)[cc]{\bf Permanent heating}}
\put(2.25,10){\makebox(0,0)[cc]{\bf by shock, {\it type IIn}}}
\put(56,16){\makebox(0,0)[cc]{\bf Fading heating}}
\put(56,10){\makebox(0,0)[cc]{\bf by radioactivity, {\it type Ia, Ib/c}}}
\end{picture}
\end{center}
Remarkably, the details of the explosion are not
important for explaining the  light curves of many supernovae that are born
from giant stars that retain their huge hydrogen envelopes.
Successful SNII light models were already constructed
four decades ago (largely by the work of the Soviet group
in Moscow and Riga \cite{GIN:1971}) due to this insensitivity to details.
The light of SNe~II is the manifestation of entropy produced  during
a short period (hours to days) of the shock propagation in the body of the presupernova star.
We cannot exclude
the possibility  that in some rare cases the SNe~II are produced by
thermonuclear explosions, not by a collapse, inside hydrogen envelopes.
This may be the case for SNe~IIn, and especially, SN2006gy.
A supernova of type~II shines for several months thanks to the
heat stored in its body (the shock dies quickly), while in an SNIIn the heat (entropy) is replenished by the shock living several months.

If the hydrogen is lost by a massive star, then we have an SNIb/c.
The same (still unknown in details) core-collapse mechanism may lead to their explosions, as in SNe~II,
however, the light is due now to entropy produced by radioactivity:
the decays   $^{56}\mbox{Ni}\to^{56}\mbox{Co}\to ^{56}\mbox{Fe}$
which lead to a slower heating of ejecta.
This way of producing light is most important for SNe~I of all subtypes.
However, some contribution of radioactivity is clearly present in late light
curves of type II supernovae as well, and for  SN1987A in LMC
it was important already before its maximum light, a month after the explosion.

If the radioactive mechanism was responsible for the light of SN2006gy, then
the amount of  $^{56}\mbox{Ni}$  must be higher than $10\, M_\odot$  \cite{Woo02,Nomoto:2007,UN:2007} .
This immediately implies a very large mass for the presupernova star, more
than $100\,M_\odot$. More important, this implies a huge explosion energy,
$(50-80)\times10^{51}$ erg.
\begin{table}
\begin{tabular}{llc}
\hline
  \tablehead{1}{l}{b}{Main Seq. Mass}
  & \tablehead{1}{l}{b}{He Core}
  & \tablehead{1}{c}{b}{Supernova Mechanism} \\
\hline
   $   10 \leq M \leq 95 $ &  $2 \leq M \leq 40$  &    Fe core collapse to a neutron star or a black hole \\
 $   95 < M \leq 130 $     &  $40 < M \leq 60$    &    Pulsational pair instability followed  by Fe core collapse \\
 $   130 < M \leq 260 $    & $60 < M \leq 137$    &     Pair instability supernova \\
  $  260 < M $             &  $ 137 < M $         &   Black hole. Possible GRB ? \\
\hline
\end{tabular}
\caption{Four kinds of deaths for non-rotating stars}
\label{tab:fate}
\end{table}
One can delineate four kinds of deaths for massive stars, see Table~\ref{tab:fate}.
There is some uncertainty about the exact values due to uncertainties in mass-loss, rotation etc.
Such  massive stars do exist, and they can have powerful explosions
in their oxygen cores which experience instability due to  the creation of large
numbers of electron-positron pairs.

\section{Pair instability supernovae}

\begin{figure}
  \resizebox{16pc}{!}{\includegraphics{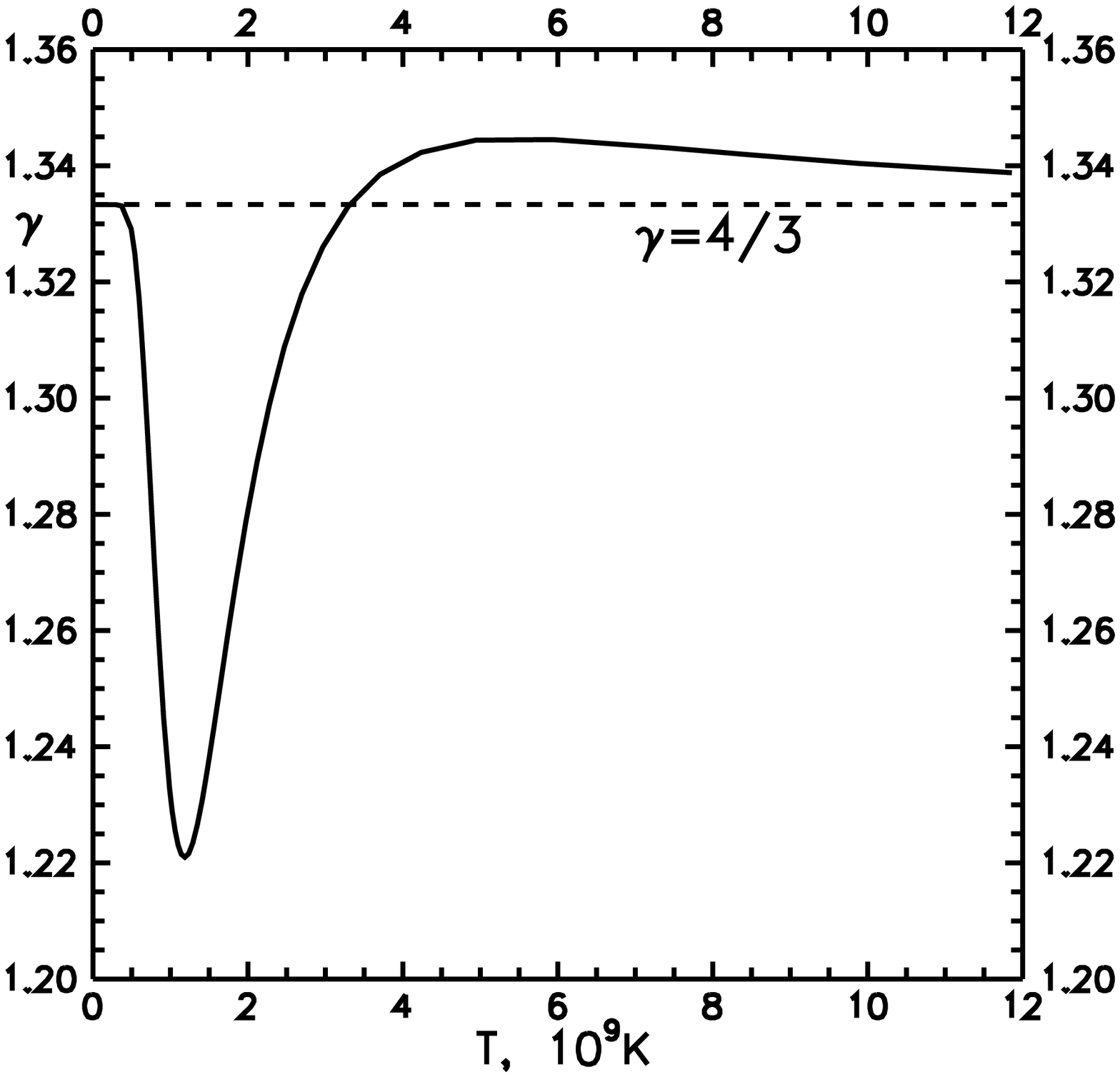}}
   \resizebox{13pc}{!}{\includegraphics{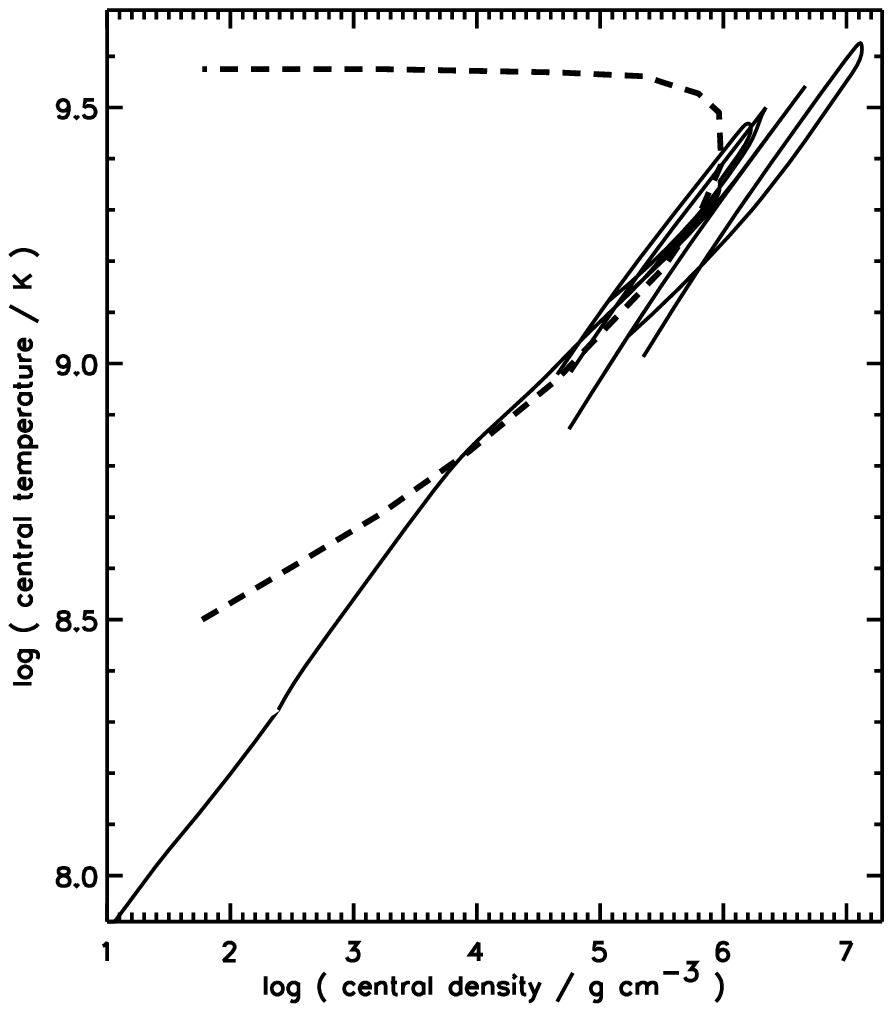}}
  \caption{Left: adiabatic exponent in the low density asymptotic limit taking into account
pair creation \cite{Nadyozhin:1974}, see also \cite{BDBN:1996}.
Right: evolution track for  one of the models \citep{WSH:2007} with initial $M \sim 103\, M_\odot$ (solid line).
 The approximate boundary of pair-instability is shown by the dashed line}
\label{dknwootc}
\end{figure}

The word `instability' refers here to hydrodynamics, to mechanical equilibrium,
not to the process of pair creation which is quite stable and reversible
 in a thermodynamic sense
in stellar interiors.
A massive star loses its mechanical stability when pairs are being created because
the adiabatic exponent $\gamma$ goes down at $T\sim 0.1$~MeV , see Fig.~\ref{dknwootc}:
the work of contraction is spent in  creating new particles and not for raising
the momenta of particles that already exist and which provide for the
equilibrium pressure.

We can easily estimate the path that leads the star into the domain of the pair-creation
instability.
From the virial theorem, for a star of mass $M$ and radius $R$, omitting all coefficients of order unity,
$ P_cV \sim P_cR^3 \sim G_{\rm N}M^2/R$.
Hence, the pressure $P_c$  in the center is
$ P_c \simeq {G_{\rm N}M^2/ R^4}$, while the density $\rho_c \simeq {M/ R^3}$,
and they are related as
$  P_c \simeq G_{\rm N}M^{2/3}\rho_c^{4/3} $.
So, if we have a classical ideal plasma with $P={\cal R}\rho T/\mu$,
where ${\cal R}$ is the universal gas constant, and $\mu$ -- mean molecular mass, we find
$$
 T_c \simeq {G_{\rm N}M^{2/3}\rho_c^{1/3} \mu / {\cal R}}.
$$
If we have a relativistic addition of $aT^4$ to $P$, the law $ T_c \propto \rho_c^{1/3} $ is the same
(but the coefficient is a bit different).
These relations have been already shown at this conference by Naoki Yoshida and by Marco Limongi.
The right panel of Fig.~\ref{dknwootc} shows how a massive star follows the law
$ T_c \propto \rho_c^{1/3} $ until entering the domain of  pair-instability.
We had already computed the light curves of pair-instability supernovae for $M>130\,M_\odot$
(third line in Table~\ref{tab:fate})
some  years ago (with S.Woosley and A.Heger), but we do not like them in the case
of SN2006gy because we do not see the evidence for a  tremendously high explosion energy in that case.
If the explosion  energy  were 2 orders of magnitude higher than for normal
SNe, then it should be seen in very broad spectral lines.
What we see in SN2006gy is different: it has narrow P Cyg lines (hence, it is type IIn)
superimposed on moderately broad emission component ($\sim 5\times 10^3$ km/s \cite{Smith}).
There is no sign of huge kinetic energy in this event.

\subsection{Multiple ejections in SNIIn}

Supernovae of type IIn  are among the most powerful transients in visible light.
No radioactive material is needed to explain their light during the first several months:
the light is produced by a long living radiative shock in a dense circumstellar envelope.
This is the main difference with standard SNe~II where the shock breaks out
into rarefied interstellar medium and disappears quickly.
Spectra and light curves of SNIIn can be explained only when the number density\
of circumstellar matter at radii of $\sim 10^{15 - 16}$~cm (where the narrow lines are formed)
is unusually high, like $10^{9 - 10}\, \mbox{cm}^{-3}$, see Fig.~\ref{94wstruct}.
This implies that a large mass (on the order of $M_\odot$ and larger)  must be ejected within years,
or even months, before the observed SN.
The first ejection may have kinetic energy appreciably lower than a standard supernova.
The slow motion of its matter explains the narrow lines of a type IIn SN.
The supernova itself is an explosion of a normal energy, but inside a dense cloud.

Paper \cite{GN:1986} was the first to suggest that an SNIIn  had a precursor, a relatively
weak explosion ejecting a large slowly moving mass.
A dramatically high mass loss  needed for  the formation of a dense envelope
shortly before SNIIn 1994W (Fig.~\ref{94wstruct})  has been derived in \cite{ChugaiBlinn}.
SNIIn 1995G is explained in \cite{cd03} in a model similar to the one  presented in \cite{ChugaiBlinn}.
Double explosions may be observed also for other SN types \cite{Pastorello:2007}.

According to \cite{ww79}   an $\sim 11\,M_{\odot}$
star might produce strong flashes in the semi-degenerate O-Ne-Mg core a few years prior
to SN explosion and the strongest flash could eject most of the
hydrogen envelope with velocities  $\sim 100$ km/s.
More recent evolutionary computations do not support the conclusion on strong
Ne flashes, but this complicated problem deserves further investigation.

\subsection{Pulsational pair-instability}
\begin{figure}
   \resizebox{14pc}{!}{\includegraphics{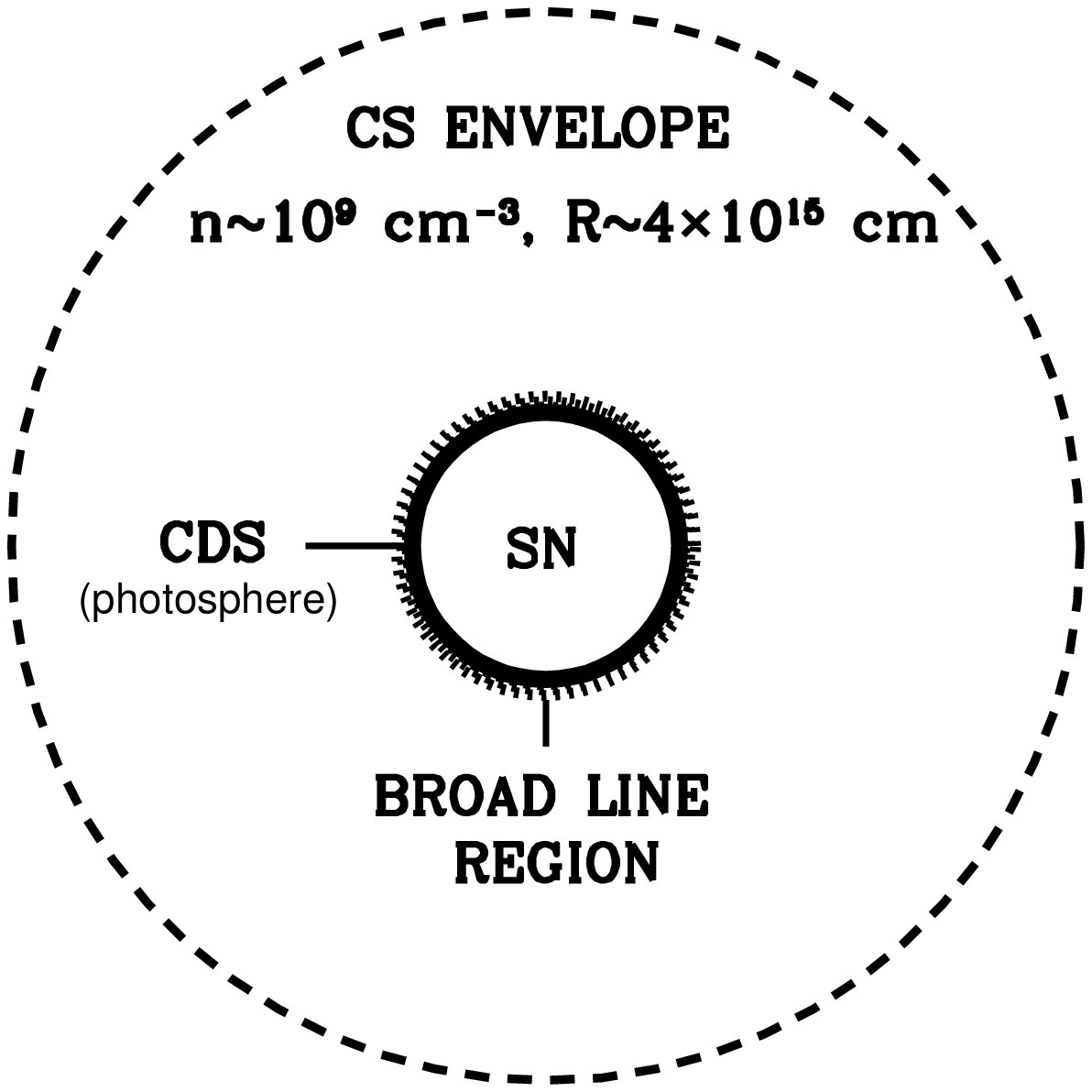}}
   \resizebox{16pc}{!}{\includegraphics{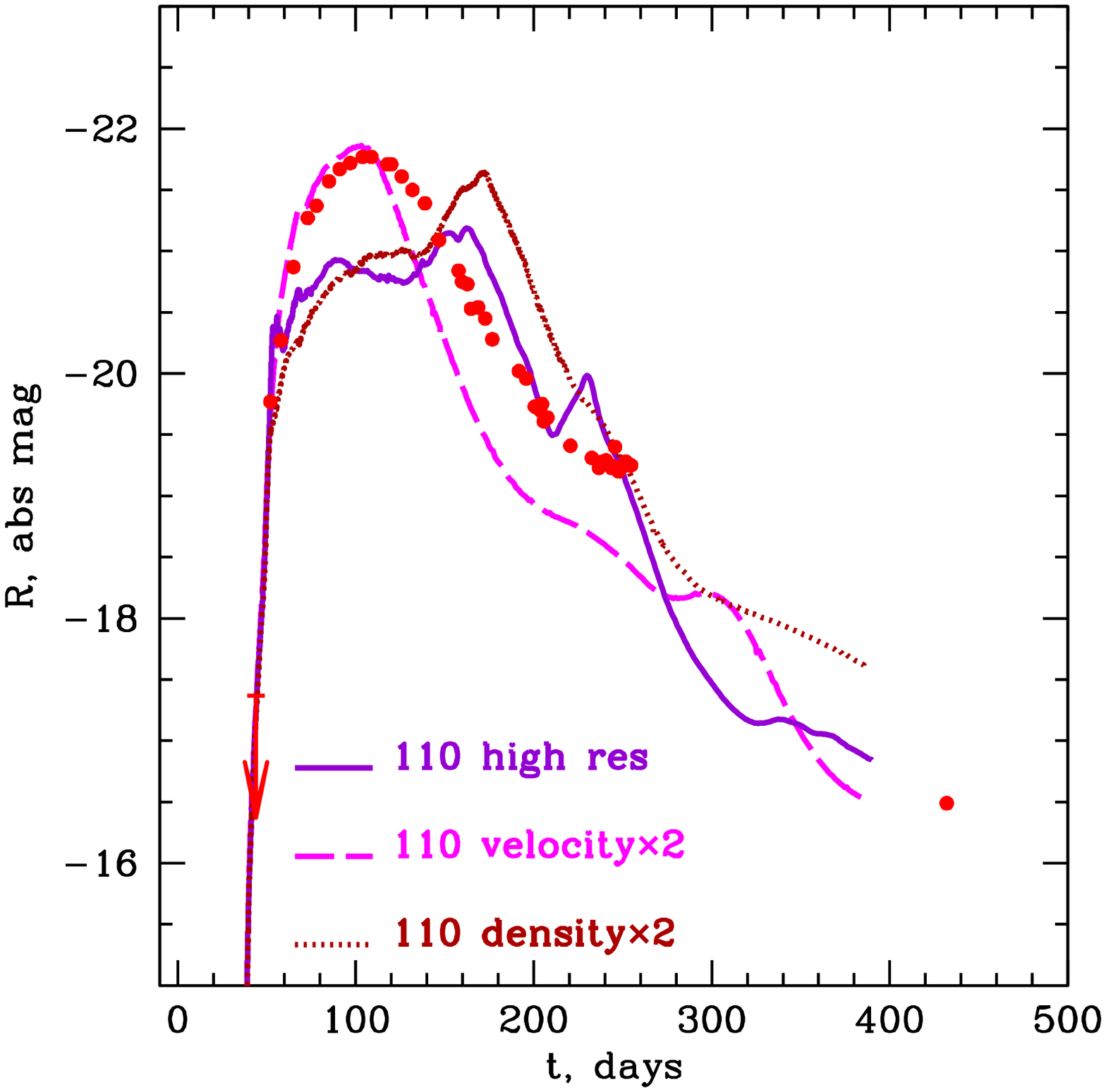}}
  \caption{Left: SN1994W structure \citep{ChugaiBlinn} prototypical for SNe~IIn.
  Right: light curve models for SN2006gy. Dots -- observations \cite{Smith},
the last observed point is from \cite{Subaru}, courtesy M.Tanaka.
The solid line is the model discussed in the text (with numerical resolution higher than in \cite{WSH:2007}),
and the dashed one where the velocity of all the ejecta has been multiplied by 2
(hence an artificial increase in the explosion energy to $2.9\times 10^{51}$ erg).
The dotted line is for the doubled density in ejecta
}
\label{94wstruct}
\end{figure}

Another workable mechanism for multiple-explosion SNe has been proposed
in our paper \cite{WSH:2007} to explain SN2006gy.
It works for a high initial mass of the presupernova star $\sim 110\, M_\odot$ (see Fig.~\ref{94wstruct}).
This mechanism is also based on the pair-creation instability,
but there is no catastrophic collapse or full explosion of the star.

The second line of Table~\ref{tab:fate} shows that  between 95 and 130 $M_\odot$
a relatively unexplored phenomenon of \emph{pulsational} pair
instability supernova \citep{Bar67,Woo86,Heg02} occurs.
An instability in the mechanical equilibrium is encountered, as in the heavier stars,
during the evolution along  $ T_c \propto \rho_c^{1/3} $ path (Fig.~\ref{dknwootc}).
A thermonuclear explosion of oxygen occurs, but the energy released is inadequate to unbind the entire star.
It suffices, however, to eject many solar masses of surface material, including  the hydrogen envelope, in a series of giant `pulses' (explosions of various strength).

The binding energy for the hydrogen envelope of $(95 - 130) M_\odot$ stars is
$\sim (0.1 - 1) \times 10^{49}$ erg while the energy of an explosive pulse is
about two orders of magnitude higher.
The velocity is in the range $100 \div 5000$~km/s depending on amount of explosive burning and mass ejected.
After a pulse, the remaining core contracts searching for a new equilibrium state.
It obeys the $ T_c \propto \rho_c^{1/3} $ law again, but now the mass is lower, so the track
is a bit different (see Fig.~\ref{dknwootc}).
The time required for the contraction is sensitive to the strength of the first pulse.
If it cools down severely after the pulse expansion, it may be
centuries before the star ignites burning again. If it remains hotter than $1.5 \times 10^9$ K, it may only take days.
After one, two or several explosion pulses the remnant of the massive star continues to live,
contrary to other supernovae, and eventually it should collapse.

To explain SN2006gy, we consider the evolution of a star with initial mass $110\,M_\odot $.
The evolution is calculated using the Kepler code \cite{Wea77,Woo02} with reduced mass loss.
The star encounters pair-instability with a total mass of  $75\,M_\odot $ (a helium core of $50\,M_\odot$)
and experiences the first pulse ejecting a cloud with mass $\approx 25\,M_\odot $ and $E_{\rm kin} \approx 1.4 \times 10^{50}$~erg.
About 7 years later the remaining core produced the second explosion with $\approx 10\,M_\odot $
and  $E_{\rm kin} \approx 7.2 \times 10^{50}$~erg.
Thus, the second ejection was faster and it had sent a shock wave into the massive cloud.
The light from this radiating shock we see as SN2006gy (Fig.~\ref{94wstruct}).
The shock is still inside the cloud, it has not yet broken out, more than one year
after the explosion of SN2006gy.

\section{Soft X-ray?}

There were arguments against the model of a shock moving into shells from previous explosions for SN2006gy.
One can estimate the temperature of the shock using standard formulas for a rarefied medium
and find that it must be very high, so SN2006gy must be a powerful source
of X-ray emission.
{\it Chandra X-ray observatory}  has measured the X-ray flux on 2006 Nov 14 from the
direction to SN2006gy. A best fit for X-ray luminosity in
(0.5--2 keV) is $1.65 \times 10^{39}$~{erg/s} \cite{Smith} , but it is a few orders
of magnitude lower than expected in the naive estimates.

\begin{figure}
  \resizebox{16pc}{!}{\includegraphics{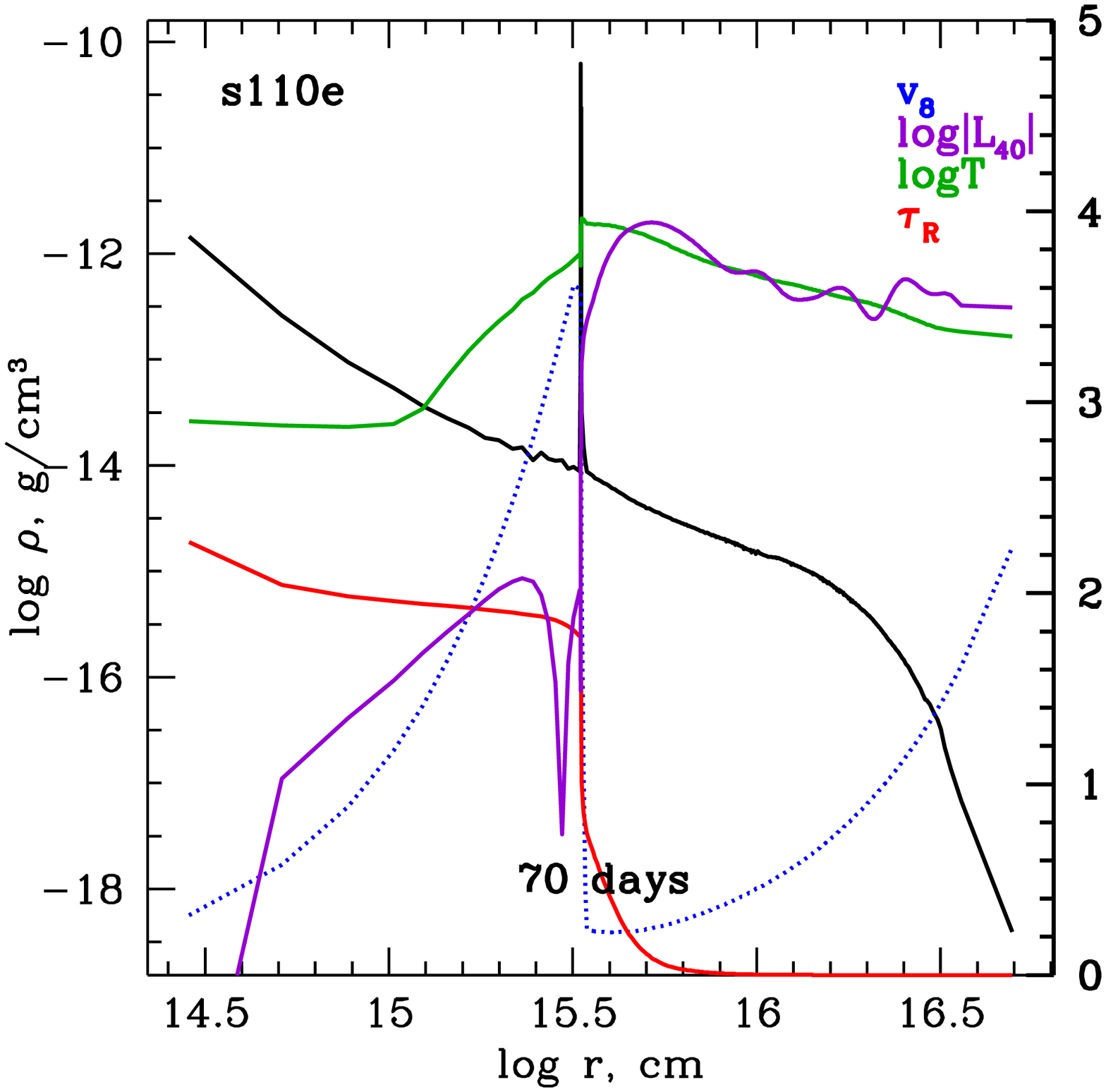}}
  \resizebox{16pc}{!}{\includegraphics{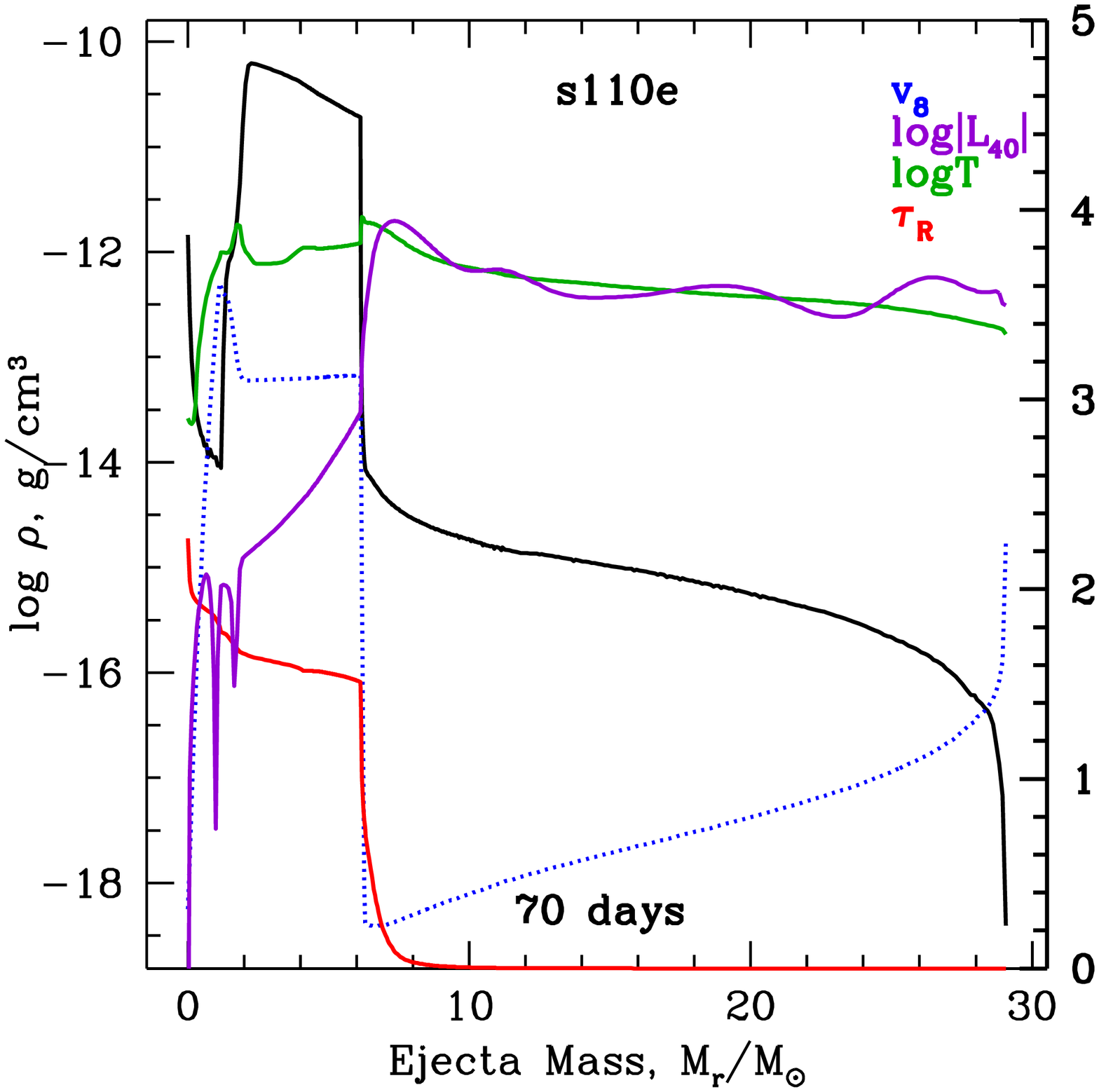}}
\caption{Hydrodynamic structure at 70 days as a function of radius (left panel) and of a Lagrangean  mass coordinate
(right panel). Black solid line is density (left Y-axis). Violet solid line is logarithm of
absolute value of luminosity $|L_{40}|$ (units $10^{40}$ erg/s).
Dotted blue line is velocity $v_8$ (units $10^8$ cm/s), green line is logarithm of temperature $T$ (in K),
and the red line is Rosseland optical depth $\tau$.
The scale for
$\log L_{40}$, $v_8$, $\log T$, and $\tau$ is on the right  Y-axis.}
\label{hyds120d}
\end{figure}

There is no problem with this in our model: the cloud of 25 solar masses is
almost transparent to the visible light of the shock because it is almost neutral,
but exactly due to this reason it is fully opaque to X-ray light.
A large mass lies above the shock, see Fig.~\ref{hyds120d}.
The zero of the mass coordinate in Fig.~\ref{hyds120d} 
is the inner edge of the ejecta.
The radiative shock is located at $M_r \approx 6 M_\odot$, or $\log r \approx 15.5$. 

If we look into other data on  SNe~IIn we see that all of them are discovered late in X-rays!
See Table~\ref{tab:xraySNiin}.
This is no wonder in our models: the shock is buried in the material of the first ejection.

\begin{table}
\begin{tabular}{llcc}
\hline
  \tablehead{1}{l}{b}{SN}
  & \tablehead{1}{l}{b}{galaxy}
  & \tablehead{1}{c}{b}{distance}
  & \tablehead{1}{c}{b}{discovery}
 \\
  \tablehead{1}{l}{b}{}
  & \tablehead{1}{l}{b}{}
  & \tablehead{1}{c}{b}{ Mpc}
  & \tablehead{1}{c}{b}{ day}
 \\
\hline
1986J   &  NGC 891   &   9.6  &    3,300 \\
1988Z  & +03-28-022  &  89   &     2,370 \\
1994W  &   NGC 4041   &        25    &      1,180 \\
1995N   &     -2-38-017    &     24 &           440 \\
1998S   &    NGC 3877  &        17  &          678 \\
2005ip  &     NGC 2096  & 30 &          490 \\
2005kd &   PGC14370   &      64  &        450 \\
\hline
\end{tabular}
\caption{Data on X-ray observations of SNe~IIn  \cite{Immler}
}
\label{tab:xraySNiin}
\end{table}

An alternate explanation for the low X-ray flux is possible.
After the work of physicists on nuclear explosions in atmosphere
in the 1940s and 1950s, we know that the preheating effect becomes so large in
strong supercritical shocks that  the {\it viscous jump} in pressure and density diminishes
and completely {\it disappears}.
In {\it radiation dominated shocks} not only the preheating effect is important; in addition
the {\it momentum transfer} from photons to electrons (and hence to ions, via the electric field) is very large.
This also destroys the viscous jump at the shock front.
Imshennik and Morozov \cite{IM:1964} have found with an accurate accounting  of the photon transfer that this happens when $P_r/P_g \simeq 8.5$.
This effect implies that postshock temperature may be so low that it does not attain keV range while inside the envelope.
All the heat is taken away by `cold'  photons.
Thus, the {\it Chandra} results \cite{Smith} tell us something about the interaction of the the first pulse ejecta
with the ISM, not about the main shock penetrating the pre-ejected shell, which shines in visible light as for SN2006gy.

\section{Conclusions}

\begin{enumerate}
 \item Radiating shocks are the most probable sources of light in most luminous
supernovae like SN2006gy.
\item {The medium in which the shining shock propagates is naturally produced
 in massive star evolution due to violent non-linear pulsations when $e^-e^+$ pairs become
appreciable in the pressure in their interiors}.

\item The supercritical radiative shock must be well below the X-ray temperature.
Even if the shock becomes hot, the overlaying matter absorbs X-rays for a long time.

\item {If SN2006gy is a pulsational-pair-instability SN, a better understanding of mass-loss is needed.
Otherwise, one has to find other evolution paths to SNe with double (or multiple) explosions.}

\end{enumerate}


\begin{theacknowledgments}
My work on SN2006gy  was supported by NASA (at UCSC), and  by the Russian Foundation for Basic Research and Science Schools (at ITEP) and by the grant IB7320-110996/1 of the Swiss National Science
Foundation. My visit to Tokyo University and to this conference is supported by RESCEU.
I am very grateful to S.~Woosley and K.~Nomoto for their warm hospitality at their Institutions and for fruitful collaboration,
to M.~Tanaka for sharing the observational data, and to L.~Rudnick for valuable comments and improvement of my English
text.
\end{theacknowledgments}

\vspace{-5mm}


\bibliographystyle{aipproc}   


\begin{thebibliography}{9}

%
%
%

\bibitem{Ofek}  E.~ Ofek, et al., \emph{ApJ} {\bf 659},  L13 (2007).

\bibitem{Smith} N.~ Smith, et al., \emph{ApJ} {\bf  666},  1116 (2007).

\bibitem{GN:1986}
E.~Grasberg,  D.~Nadyozhin,  \emph{Soviet Astronomy Letters}  \textbf {12},  68 (1986).

\bibitem{ChugaiBlinn}
N.~Chugai, S.~Blinnikov et al., \emph{MNRAS}  \textbf{352},  1213 (2004).

\bibitem{GIN:1971} E.~Grasberg, V.~Imshennik, D.~Nadyozhin, D.~K.  
\emph{Astrophysics and Space Science} \textbf{10}, 28-51 (1971).

\bibitem{Woo02}
S.~Woosley, A.~Heger, and T.~Weaver,
\emph{ Rev.Mod.Phys.}  \textbf{74}, 1015 - 1071 (2002).

\bibitem{Nomoto:2007} K.~Nomoto,  et al., {\it American
Institute of Physics Conference Series} {\bf 937}, 412 (2007).

\bibitem{UN:2007} H.~Umeda, and K.~Nomoto,
{\it ArXiv e-prints} {\bf 707}, arXiv:0707.2598 (2007).

\bibitem{Nadyozhin:1974} D.~Nadyozhin, {\it Nauchnye Informatsii} {\bf 32}, 3 (1974).

\bibitem{BDBN:1996}S.~Blinnikov, N.~Dunina-Barkovskaya,  and D.~Nadyozhin,   {\it ApJS} {\bf 106}, 171
    (1996).

\bibitem{WSH:2007}
  S.~Woosley, S.~Blinnikov, A.~Heger,  {\it Nature} {\bf 450}, 390  (2007).

\bibitem{cd03} N.~Chugai and I.~Danziger,  \emph{Astr.\ Lett.}
  \textbf {29}, 732 ( 2003)

\bibitem{Pastorello:2007} A.~Pastorello,  et al., {\it Nature} {\bf 447}, 829 ( 2007).

\bibitem{ww79}  T.~Weaver and S.~Woosley,  \emph{BAAS} \textbf {11}, 724 (1979).

\bibitem{Bar67} Z.~Barkat,  G.~Rakavy, and N.Sack,
 \emph{Phys.Rev.Lett.} \textbf {18}, 379 - 381 (1967).

\bibitem{Woo86}
S.~Woosley  and T.~Weaver, ``The Physics of Supernovae'', in
 \emph{Radiation Hydrodynamics in Stars and Compact Objects}, IAU Colloq.~89 Proceedings,
edited by D.~Mihalas and K.-H.~Winkler, Springer-Verlag,  Lecture Notes in Physics 255, 1986, pp.  91 - 120.

\bibitem{Heg02}
A.~Heger and S.~Woosley,  
 \emph{ApJ}  \textbf{567}, 532 - 543 (2002).

\bibitem{Wea77}
T.~Weaver, G.~Zimmerman, and S.Woosley,
  \emph{ApJ}  \textbf{225}, 1021 - 1029 (1978).

\bibitem{IM:1964}
V.~Imshennik and Yu.~Morozov,
\emph{Zhurnal Prikladnoj Mekhaniki i Tekhnicheskoj Fiziki} \textbf{No. 2}, 8-21 (1964).

\bibitem{Immler}
S.~Immler,  \url{http://lheawww.gsfc.nasa.gov/users/immler/} (2007).

\bibitem{Subaru}
K.~Kawabata, M.Tanaka, et al. (2008), submitted to   \emph{ApJ}.

\end{thebibliography}

%


\end{document}